\documentclass[twocolumn]{aastex62}

\usepackage{amsmath}
\shorttitle{$\eta_\Earth$ estimates}
\shortauthors{Pascucci et al.}

\begin{document}

\title{The impact of stripped cores on the frequency of Earth-size planets in the habitable zone}

\correspondingauthor{Ilaria Pascucci}
\email{pascucci@lpl.arizona.edu}

\author{Ilaria Pascucci}
\affiliation{Lunar and Planetary Laboratory, The University of Arizona, Tucson, AZ 85721, USA}
\affiliation{Earths in Other Solar Systems Team, NASA Nexus for Exoplanet System Science}

\author{Gijs D. Mulders}
\affiliation{Department of Geophysical Sciences, The University of Chicago}
\affiliation{Earths in Other Solar Systems Team, NASA Nexus for Exoplanet System Science}

\author{Eric Lopez}
\affiliation{NASA Goddard Space Flight Center, 8800 Greenbelt Rd, Greenbelt, MD 20771, USA}



\begin{abstract}
The frequency of Earth-size planets in the habitable zone of Sun-like stars, hereafter $\eta_\Earth$,  is a key parameter to evaluate  the yield of nearby Earth analogues that can be detected and characterized by future missions. Yet, this value is poorly constrained as there are no reliable exoplanet candidates in the habitable zone of Sun-like stars in the {\it Kepler} field. Here, we show that extrapolations relying on the population of small ($< 1.8\,R_\oplus$) short-period ($< 25\,$days) planets bias  $\eta_\Earth$ to large values. 
As the radius distribution at short orbital periods is strongly affected by atmospheric loss, we re-evaluate $\eta_\Earth$ using exoplanets at larger separations.
We find that $\eta_\Earth$ drops considerably, to values of only $\sim 5-10$\%. Observations of young ($< 100$\,Myr) clusters can probe short-period sub-Neptunes  that still retain most of their envelope mass. As such, they can be used to quantify the contamination of sub-Neptunes to the population of {\it Kepler} short-period small planets and aid in more reliable estimates of $\eta_\Earth$.
\end{abstract}

\keywords{methods: data analysis --- planets and satellites: detection --- planets and satellites: terrestrial planets --- surveys}


\section{Introduction} \label{sec:intro}
The past decade has seen an exponential increase in the number of known exoplanets, mainly thanks to the  NASA’s {\it Kepler} space telescope (e.g., \citealt{borucki11,borucki17}). One of the most interesting and surprising results from this mission has been the discovery of a multitude of short-period planets (e.g., \citealt{fressin13,petigura13}), located much closer to their star than Mercury to the Sun. Follow-up observations of a subset of these planets \citep{johnson17,petigura17} led to more precise stellar, hence planetary, radii and to the discovery of  the so-called radius valley, a much lower frequency of planets with radii $\sim 1.8\,R_\oplus$ rather than $\sim 1.3\,R_\oplus$ or $\sim 2.4 \,R_\oplus$  \citep{fulton17}. Using the sub-set of exoplanet host stars with parameters homogeneously measured from asteroseismology,  \citet{vaneylen18}  confirmed the presence of the radius valley. Furthermore, they reported that the valley has a weak inverse dependence with orbital period as $\propto P^{-0.09}$, which has been recently confirmed by \citet{martinez2019}.

What is the origin of the radius valley? \citet{owen13} predicted early on that photoevaporation driven by high-energy stellar photons could herd planet radii into a bi-modal distribution, closely matching that subsequently found by \citet{fulton17}. 
This happens because photoevaporation is least efficient for planets that have twice the core radius, or an H/He-rich envelope that is just a few \% of the total mass: lighter or more massive envelopes are unstable and by evaporating efficiently end up populating one of the two peaks of the planet radius distribution
(see Figure~6 in \citealt{owenwu17} but also \citealt{lopezf13,jinmordasini18,lopezrice18}). Alternatively, \citet{ginzburg18} suggested that the cooling luminosity of the planet itself drives atmospheric loss: light atmospheres, where the ratio between the heat capacity of the core and the envelope is $\leq$5\%, are mostly heated by the underlying rocky core and are rapidly removed while more massive atmospheres regulate their own cooling and can survive. 

Importantly,  both scenarios imply that the population of short-period ($<100\,$days) small ($< 1.8\,R_\oplus$) planets is  contaminated by sub-Neptunes that have lost a significant fraction of their envelope mass.
Unlike Earth, these planets 
formed within few Myr in a gaseous circumstellar disk from which they accreted their envelope (e.g., \citealt{lee16}).  This conclusion is further corroborated by the expectation that a primordial rocky population, born after disk dispersal, should result in a larger planet mass, hence radius, with increasing semi-major axis (e.g., \citealt{lopezrice18}) which is opposite to the observed radius valley dependence with orbital period. 

As the {\it Kepler} exoplanet detectability decreases rather steeply toward small planet radii and large orbital periods  and no true Earth analog\footnote{With Earth analog we mean Earth-size planet with an orbital period of 1 year.} has been discovered around Sun-like stars  (e.g., \citealt{burke15,borucki17,thompson18}), $\eta_\Earth$ cannot be directly measured. Values obtained from M or K dwarfs (e.g., \citealt{dc15}) likely provide an upper limit as  small planets are more common around low-mass stars (e.g., \citealt{mulders18_book} for a recent review on planet populations as a function of stellar properties). For Sun-like G-type stars, planets with either larger radii or much closer in to their stars  have become crucial to estimate the frequency of Earth-size planets in the Habitable Zone  (HZ), hereafter $\eta_\Earth$,  see also Section~\ref{sec:estimate}. 
\citet{lopezrice18} pointed out that fitting separable power laws in planet radius and period will likely lead to overestimate  $\eta_\Earth$ as the 
radius distribution will be dominated by short-period planets, many of which could be stripped cores, while the period distribution will be dominated by non-rocky sub-Neptunes.

Here, we begin to evaluate the impact of  short-period planets on $\eta_\Earth$ in a systematic way. First, we explain our definition of the HZ and review the methods and $\eta_\Earth$  estimates reported in the literature (Section~\ref{sec:estimate}). Next, we adopt the  latest {\it Kepler} DR25 catalogue \citep{thompson18} with stellar properties from Gaia DR2 \citep{berger18}, in combination with the Exoplanet Population Observation Simulator \texttt{epos}\footnote{Here we use the \texttt{epos} version 1.1.0 retrievable via https://github.com/GijsMulders/epos} (\citealt{mulders18},  hereafter M18), to evaluate the impact of  short-period planets on estimates of $\eta_\Earth$. We show that $\eta_\oplus$ drops by  factors of $\sim$4-8 when extrapolations exclude short-period planets, many of which could be stripped cores  (Section~\ref{sec:nostripped}).  As  $\eta_\Earth$ directly impacts the yield of Earth analogues that can be detected by future missions like LUVOIR and HabEx (e.g., \citealt{stark15}), it is crucial to better constrain its value. A discussion of how this could be achieved is provided in Section~\ref{sec:summary}.

\section{The occurrence of Earth-size planets in the habitable zone} \label{sec:estimate}
One of the primary science goals of the {\it Kepler} mission was to measure the frequency of Earth-size and larger planets in the HZ of Sun-like stars \citep{borucki2003}. As no true Earth analog has been detected, $\eta_\oplus$ estimates necessarily rely on assumptions based on the more abundant population of  short-period and larger planets. 

Numerous estimates of $\eta_\oplus$ are available in the literature and the ExoPAG Study Analysis Groups~13 has recently summarized and tried to reconcile discrepancies among different studies\footnote{https://exoplanets.nasa.gov/exep/exopag/sag/\#sag13}. To cancel out dependencies on the definition of the HZ and planet size range, the report focuses on comparing $\Gamma_\oplus = \frac{\partial^2 N(R,P)}{\partial {\rm ln} R \partial {\rm ln} P}|_{R=R_\oplus,P=1yr}$, i.e. $\eta_\oplus$  per log period and radius bin.
Even with this definition, literature $\Gamma_\oplus$ span more than an order of magnitude in range, from 2\% \citep{fk14} to 70\% \citep{traub15}.
The report highlights that major differences  are introduced by the use of different {\it Kepler} catalogues and completeness curves, with more recent ones giving systematically larger values, while different methods/extrapolations introduce only a factor of two uncertainty. We will show,  instead, that extrapolations are very sensitive to the exclusion of short-period planets  (Section~\ref{sec:nostripped}). Finally, the report provides power law fits in period and radius for small ($< 3.4\,R_\oplus$) planets based on the average  of 12 community occurrence rate grids which include up to the \texttt{DR24} {\it Kepler} data release. These fits imply  $\Gamma_\oplus =38\%$ or $\eta_\oplus \sim 20\%$ when considering a conservative HZ (0.95-1.67\,au) and habitable planets $\sim 0.8-1.4\,R_\oplus$, very close to the 24\% baseline value used to estimate the exoplanet yield for the LUVOIR\footnote{https://asd.gsfc.nasa.gov/luvoir/resources/docs/\\
LUVOIR\_Interim\_Report\_Final.pdf} and HabEX\footnote{https://www.jpl.nasa.gov/habex/pdf/HabEx\_Interim\_Report.pdf} mission concept studies.
 
 \begin{figure}[ht!]
   \centering
    \includegraphics[scale=0.45]{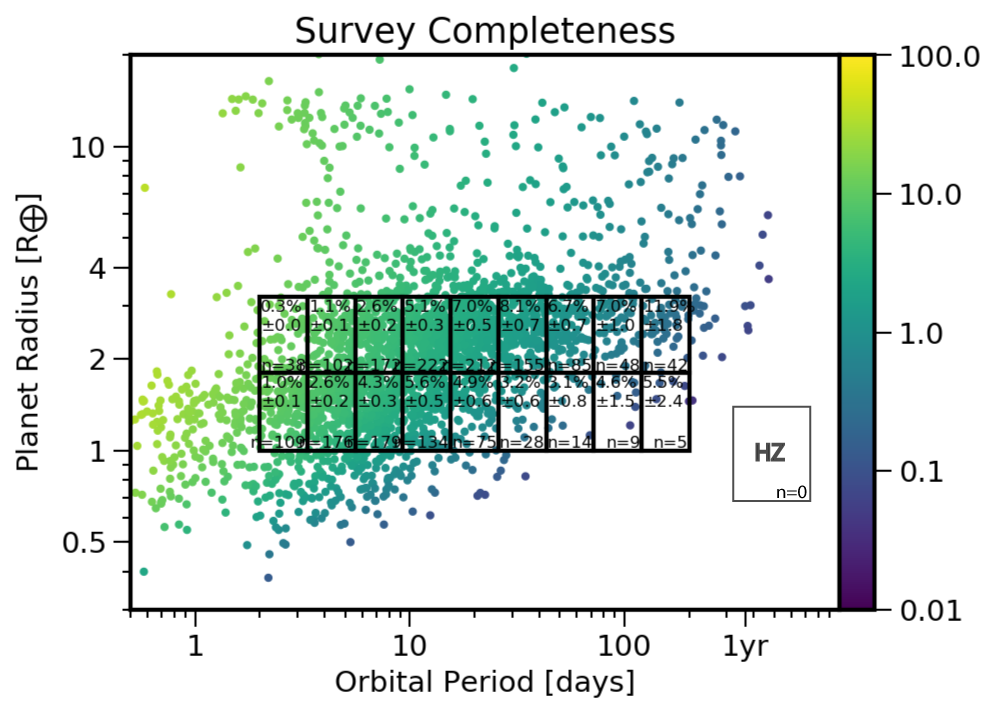}    
\hbox{\hspace{-2.5em} \includegraphics[scale=0.4]{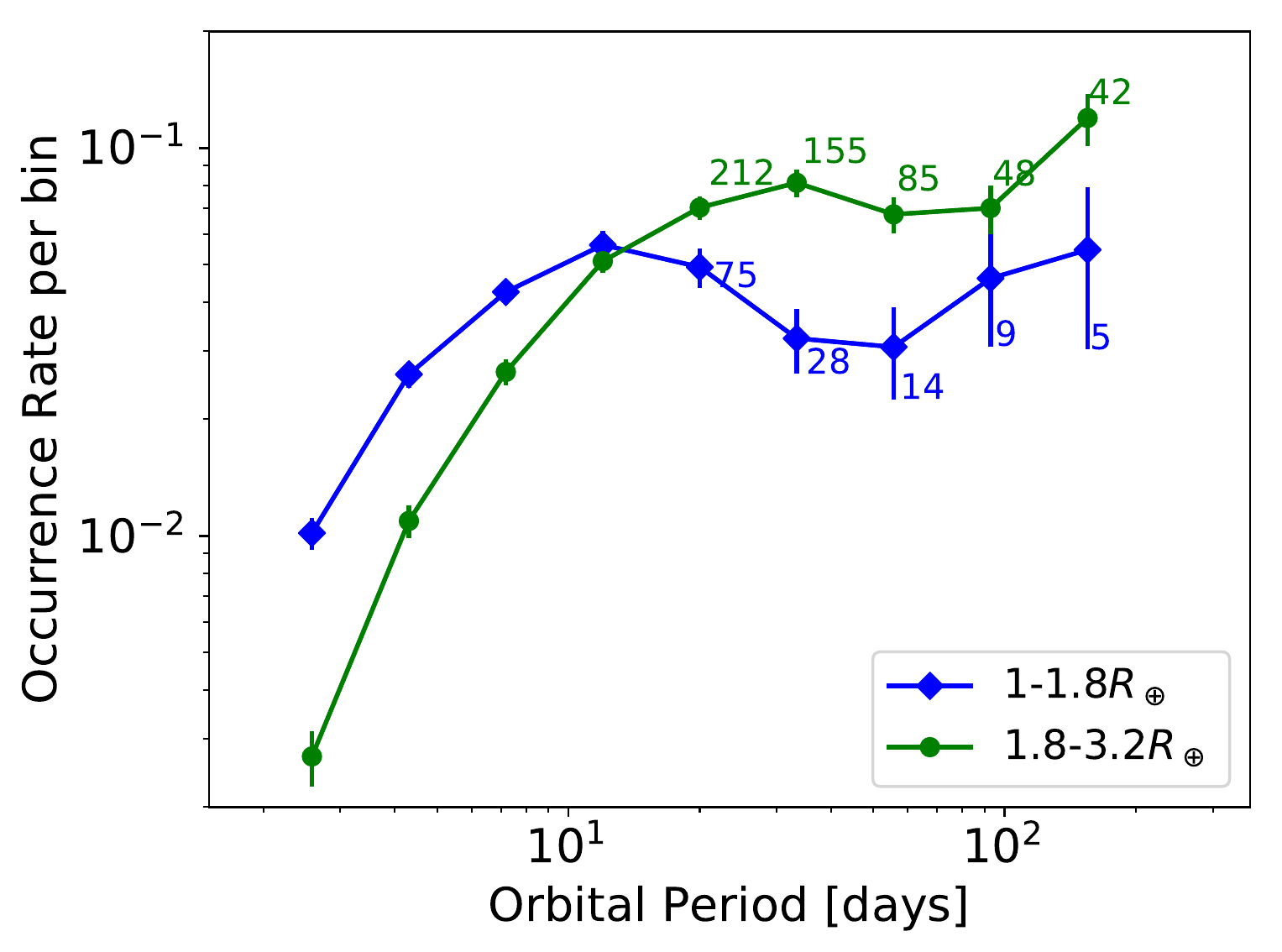}}
     \caption{Upper panel: DR25+Gaia candidate list, color coded by survey completeness. The sample includes only dwarfs and planet candidates with a {\it Robovetter} score $\ge$ 0.9. The grey rectangle delineates the HZ, no reliable planet candidate is detected inside the HZ.
 Occurrence rates using the inverse detection efficiency method are also provided for nine period and two radius bins (black rectangles). For  clarity these values are also plotted in the lower panel with the number of planets per bin for the five bins at largest orbital periods.}\label{fig:bins}
\end{figure}

In M18 we used the latest Q1-Q17 \texttt{DR25} {\it Kepler} catalogue \citep{thompson18} to present a new code, \texttt{epos}, which is based on a forward modeling approach to constrain exoplanet populations. The code includes the most recent detection and vetting efficiency curves for the most reliable planet candidates ({\it Robovetter} score $\ge$0.9). We fit two broken power laws, one in orbital period (for $2 < P < 400$\,days) and one in planet radius (for $0.5 < R < 6\, R_\oplus$),  and demonstrated that they provide a good match to the bulk of the {\it Kepler} planet candidates (see Appendix~\ref{app} for a detailed description of the equations employed in \texttt{epos}). Integrating the posterior distribution in the $0.9 < P < 2.2\,P_\oplus$ and $0.7 < R < 1.5\, R_\oplus$ range, which is based on the 
\citet{kopparapu13} conservative HZ for the most common Sun-like stars in the {\it Kepler} sample, we found $\eta_\oplus = 36^{+14}_{-14}\%$ and $\Gamma_\oplus =53^{+20}_{-21}\%$. These values agree with the baseline $\Gamma_\oplus =60\%$ obtained by \citet{burke15} by fitting a single power law in period (for $50 < P < 300$\,days) and broken power-law in radius (for $0.75< R < 2.5\,R_\oplus$) on all Q1-Q16 {\it Kepler} planetary candidates, i.e. no reliability cut was applied. 
\citet{hsu19} recently estimated $\Gamma_\oplus$ from the DR25+Gaia~DR2 catalog using a Bayesian framework and derived a  median (50th percentile) value of 57\% with 1$\sigma$ boundaries of 34\% and 84\%,  though they also did not include planets' reliability. Finally, \citet{zink19} used the same DR25+Gaia~DR2 catalogue but adopted two independent broken power-law relations as in M18 and derived essentially the same $\eta_\oplus$, 34\%,  although with a much lower uncertainty of only 2\% as they include several system's multiplicity parameters and priors to eliminate unphysical solutions.

\subsection{A  fourfold to eightfold drop in $\eta_\oplus$}\label{sec:nostripped}
To evaluate the impact of  short-period planets on $\eta_\oplus$ estimates, we adopt the same definition of HZ and habitable planets as in M18.  We update  \texttt{epos} to include Gaia-revised stellar radii for the {\it Kepler} sample \citep{berger18} and re-calculate  detection efficiency contours for each individual star using \texttt{KeplerPORTs} \citep{bc17}.  After removing giant and sub-giants as in \citet{berger18}, we obtain a sample of 119,220 dwarfs with a median mass of 0.976\,$M_\sun$. We calculate the average survey detection efficiency for this sample as well as re-compute vetting efficiency curves for the reliable ({\it Robovetter} score $\ge$0.9) candidates in our sample following the approach described in M18. 



\begin{deluxetable}{lccc}[b!]
\tablecaption{\texttt{epos} best fit solutions with 1-$\sigma$ confidence intervals \label{tab:bpar}}
\tablecolumns{3}
\tablewidth{0pt}
\tablehead{
\colhead{Parameter} &
\colhead{M18} &
\colhead{Model\#1} &
\colhead{Model\#4} 
}
\startdata
$\eta$ &4.9$^{+1.3}_{-1.2}$ & 4.6$^{+1.0}_{-1.1}$ &  2.7$^{+0.5}_{-0.3}$ \\
$P_{\rm break}$ (days) &  12$^{+5}_{-3}$ & 11$^{+6}_{-3}$ & $-$ \\
$a_{P}$ &  1.5$^{+0.5}_{-0.3}$ & 1.6$^{+0.6}_{-0.3}$ & $-$ \\ 
$b_P$ & 0.3$^{+0.1}_{-0.2}$  & 0.3$^{+0.1}_{-0.2}$  & 0.14$^{+0.07}_{-0.07}$ \\ 
$R_{\rm break}$ ($R_\oplus$) & 3.3$^{+0.3}_{-0.4}$ & 3.4$^{+0.2}_{-0.3}$ & 3.2$^{+0.2}_{-0.3}$ \\ 
$a_R$ & -0.5$^{+0.2}_{-0.2}$ & -0.3$^{+0.2}_{-0.2}$ & 1.0$^{+0.5}_{-0.5}$ \\
$b_R$ & -6$^{+2}_{-3}$ & -7$^{+2}_{-2}$ &  -6$^{+2}_{-2}$ \\
\enddata
\tablecomments{The equations used in the fit and an explanation for each of the parameters listed here are provided in Appendix~\ref{app}. Posterior distributions for Model\#1 and \#4 are shown in Figures~\ref{fig:PMulders} and \ref{fig:P12d}, respectively.}
\end{deluxetable}

Figure~\ref{fig:bins} shows our {\it Kepler} DR25+Gaia candidate list color coded by survey completeness. The grey rectangle delineates the HZ, no reliable detection is present in the region. Accepting all planet candidates, regardless of their {\it Robovetter} score, results in 4 detections, with 2 at the upper border of the box, while using pre-Gaia stellar parameters would further increase the number to 11, see also Figure~14 in \citet{burke15} for the same number of planetary candidates in the HZ from the Q1-Q16
{\it Kepler}  catalogue and pre-Gaia radii\footnote{Note that the five candidates at the completeness level of 0.01\% were already marked as suspected false positive in the SAG13 report and excluded from occurrence rate calculations}.

\begin{figure}[h!]
\minipage{0.45\textwidth}
   \includegraphics[width=\linewidth]{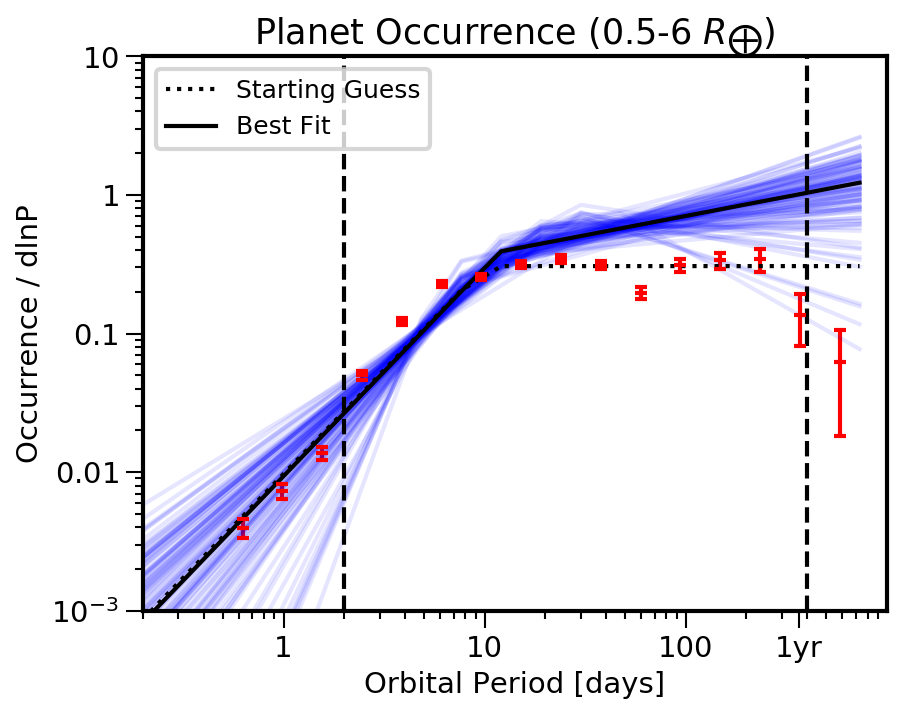}
\endminipage\hfill
\minipage{0.45\textwidth}
   \includegraphics[width=\linewidth]{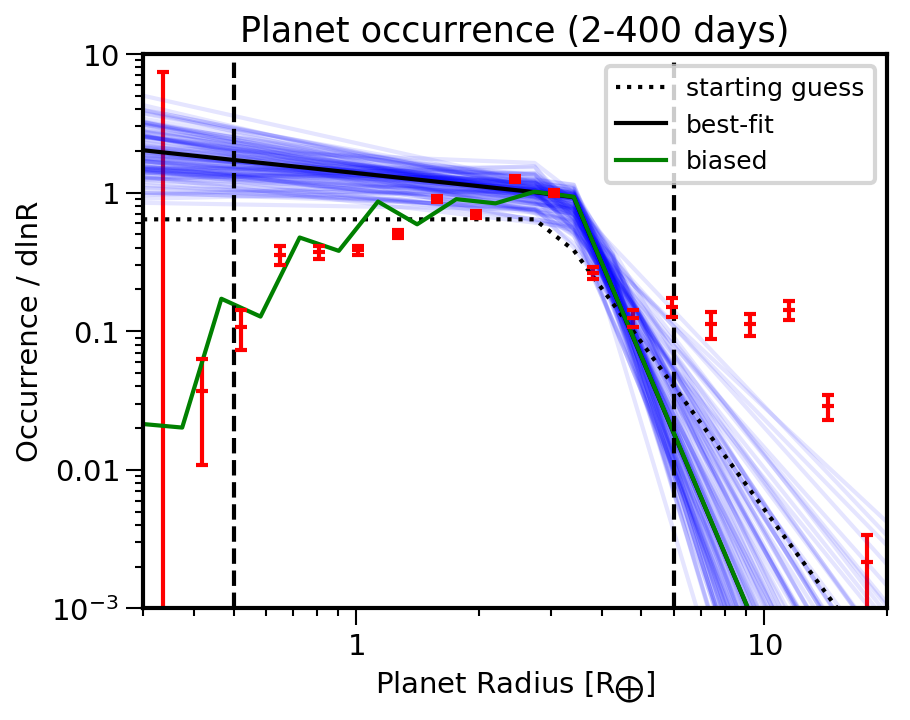}
\endminipage
\caption{\texttt{epos} posterior orbital period (top) and planet radius distributions (bottom) for Model\#1.  Black vertical dashed lines indicate the range in planet period and radius that \texttt{epos} fits.  Red points with errorbars show the occurrence rates calculated with the inverse detection efficiency method. A biased version of the posterior planet radius distribution, assuming no planets below a completeness of 0.03\%, is shown in green. The good agreement between the green curve and the red points illustrates that occurrence rates, estimated with the inverse detection efficiency method, underestimate the true distribution in bins where the completeness is low and planets are only partially detected.} \label{fig:PMulders}
\end{figure}

We first run \texttt{epos} in its Monte Carlo mode with this new {\it Kepler} DR25+Gaia catalogue, the updated completeness and vetting efficiencies, and over the same period ($2 < P < 400$\,days) and planet radius ($0.5 < R < 6\, R_\oplus$) range as in M18.  This first run is referred to as Model\#1.
We find the same best fit parameters  as M18 within the quoted 1-$\sigma$ confidence intervals, see Table~\ref{tab:bpar}. The posterior distributions (blue lines) and best-fit relations (black lines) in orbital period and planet radius  for our Model\#1 are shown in Figure~\ref{fig:PMulders}.   The same figure provides the occurrence rates calculated with the inverse detection efficiency method (red points with errorbars). As already pointed out in M18, the low values for large orbital periods ($P \ge 30$\,days) and small planet radii ($R \leq 1.5\,R_\oplus$) are just due to the inclusion of bins where the completeness is low and {\it Kepler} has only partly detected planets (see Figure~\ref{fig:bins}). To illustrate this point the green line in the lower panel of Figure~\ref{fig:PMulders} gives the biased posterior, i.e. the posterior distribution assuming that no planets are detected below a completeness of 0.03\%. The good agreement between the red points and the green line demonstrates how the classic inverse detection efficiency method can underestimate true rates (see also \citealt{fk14} and Appendix~\ref{app} for posteriors and occurrence rates over a restricted period and radius range with higher completeness). Note that the mis-match at $R > 6\, R_\oplus$ is inconsequential to the paper since the fit and the $\eta_\oplus$  calculation ignore that part of parameter space.

The key features of the best-fit relations shown in Figure~\ref{fig:PMulders} and relevant to this investigation are: i) a slight increase in the occurrence vs orbital period beyond $P_{\rm break}$ and ii) an increase in the occurrence of planets smaller than $R_{\rm break}$. By integrating the posterior distribution in the HZ we find $\Gamma_\oplus =$60$^{+22}_{-25}$ \% and  $\eta_\oplus =$41$^{+15}_{-17}$ \%, see also Table~\ref{tab:res}, the same as those reported in M18.  Expanding upon and corroborating M18, this test also shows that the Gaia-revised stellar radii have very little impact on this type of modeling, in spite of reducing by more than half the number of all candidates falling in the HZ, see also  \citet{zink19}.

Next, we run  \texttt{epos} over the same period range but only on the sample of small planets ($0.5 < R < 2\, R_\oplus$), i.e. we employ only a broken power-law in period to fit the observed distribution (Model\#2 in Table~\ref{tab:res}). This model returns similarly large  $\Gamma_\oplus$ and  $\eta_\oplus$ as Model\#1 because the  planet distribution increases toward small radii ($a_R = -1.25\pm0.05$). Finally, we run a set of models where we exclude the population of  short-period planets but fit a large range of planet radii with a broken power law. The minimum period of 12\,days (Models\#3 and \#4) is chosen to exclude the known orbital period break for sub-Neptunes \citep{howard12} while for periods $>$\,25\,days (Models\#5, \#6, and \#7) theoretical models predict minimal photoevaporation (e.g., \citealt{owenwu17}), hence negligible contamination of stripped cores. We find that  $\Gamma_\oplus$ and  $\eta_\oplus$ drop by factors of  $\sim 4-8$ when excluding the population of  short-period planets and that the results are robust against the lower planet radius boundary that is adopted. Note that Model\#7, single power laws for small planets with minimal photoevaporation, essentially provides the same $\eta_\oplus$ estimates as Model\#6  where the inclusion of large planets is modeled via a broken power-law relation in planet radius.

\begin{deluxetable}{cccccc}[t!]
\tablecaption{\texttt{epos} modeling results \label{tab:res}}
\tablecolumns{6}
\tablewidth{0pt}
\tablehead{
\colhead{Model} &
\colhead{Fitted $P$} &
\colhead{Fitted $R$} &
\colhead{Function} &
\colhead{$\Gamma_\oplus$} & 
\colhead{$\eta_\oplus$}\\
\colhead{\#} &
\colhead{days} &
\colhead{R$_\oplus$} &
\colhead{} &
\colhead{\%} &
\colhead{\%}
}
\startdata
1 & 2--400 & 0.5-6 & 2D broken & 59.6$^{+21.8}_{-25.4}$ & 40.6$^{+14.9}_{-17.3}$  \\
2 & 2--400 & 0.5-2 & $P$ broken & 78.7$^{+43.5}_{-39.2}$ & 53.6$^{+29.7}_{-29.7}$  \\
3 & 12-400 & 0.5-6 & $R$ broken & 17.0$^{+7.6}_{-5.6}$ & 11.5$^{+5.2}_{-3.8}$  \\ 
4 & 12-400 & 1-6   & $R$ broken & 16.0$^{+8.0}_{-5.5}$ &10.9$^{+5.5}_{-3.7}$  \\ 
5 & 25-400 & 0.5-6   & $R$ broken & 8.6$^{+8.9}_{-5.1}$ &5.9$^{+6.0}_{-3.5}$  \\ 
6 & 25-400 & 1-6   & $R$ broken & 8.0$^{+10.3}_{-5.4}$ & 5.4$^{+7.0}_{-3.7}$\\ 
7 & 25-400 & 1-2   &  $P \& R$ single  & 7.8$^{+10.3}_{-3.8}$ &  5.3$^{+7.0}_{-2.6}$\\ 
\enddata
\tablecomments{'2D broken' stands for broken power law in period and radius while 'P (R) broken' means that we have employed a broken power law in period (radius) and a single power law in radius (period), see Appendix~\ref{app} for the equations.}
\end{deluxetable}

\begin{figure}[h!]
\minipage{0.45\textwidth}
   \includegraphics[width=\linewidth]{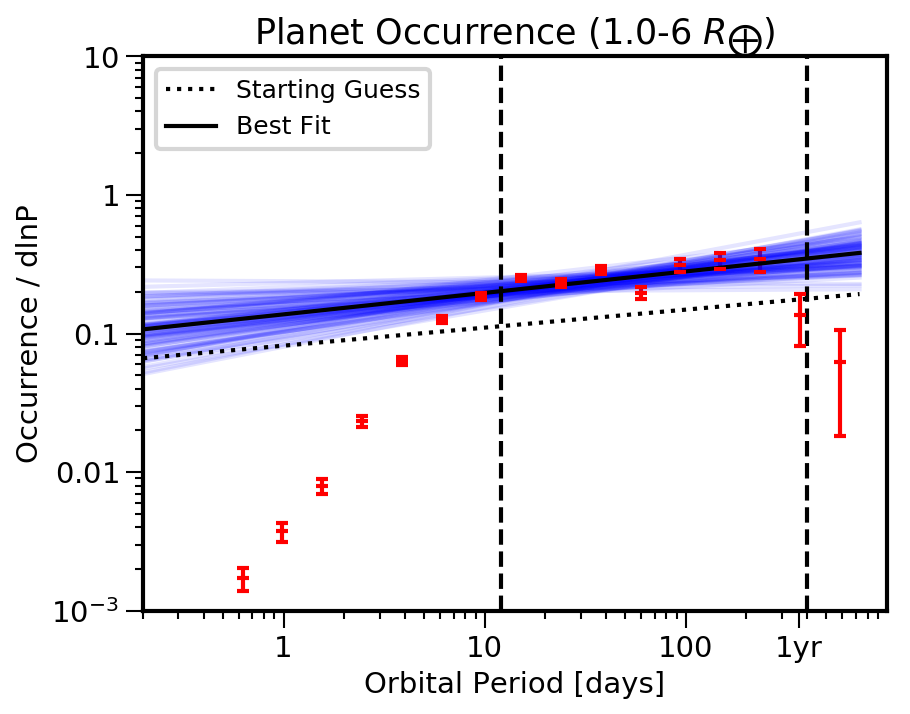}
\endminipage\hfill
\minipage{0.45\textwidth}
   \includegraphics[width=\linewidth]{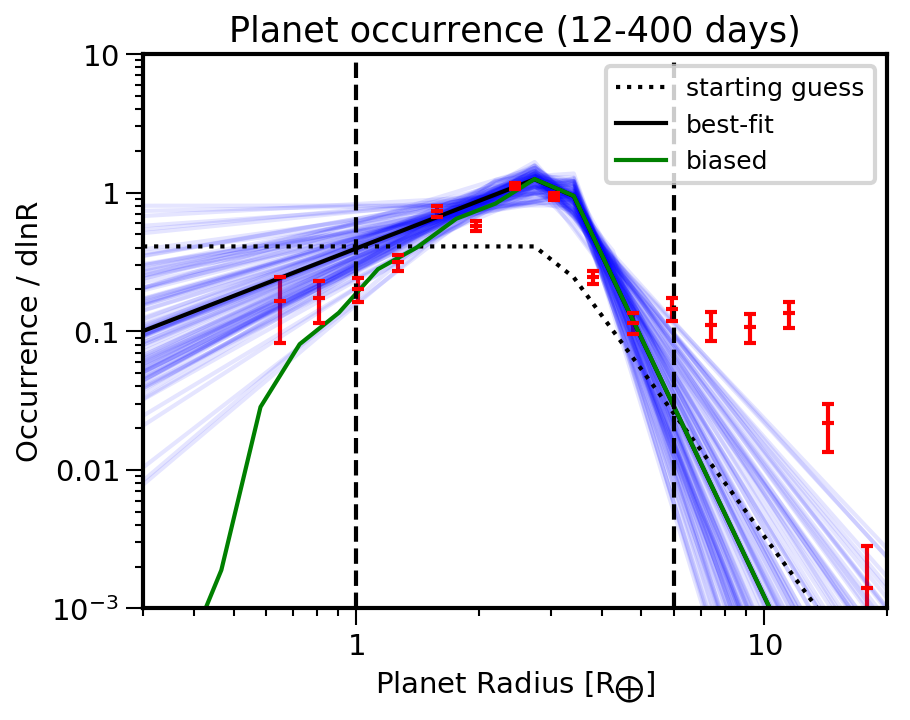}
\endminipage
\caption{\texttt{epos} posterior orbital period distribution (top) and planet radius distribution (bottom) for Model\#4.  Symbols as in Figure~\ref{fig:PMulders}. Note that these occurrence rates (red points with errorbars) are calculated for larger planets than in Figure~\ref{fig:PMulders}, hence they underestimate less the true distribution.} \label{fig:P12d}
\end{figure}

To clarify why there is such a difference in the  $\eta_\oplus$ estimates, we provide the  \texttt{epos} best-fit values and 1-$\sigma$ confidence intervals for Model\#4 in Table~\ref{tab:bpar} and the posterior orbital period and planet radius distributions in Figure~\ref{fig:P12d}. The  planet distribution is still slightly increasing toward larger orbital periods (parameter $b_P$) but steeply drops toward small planet radii (parameter $a_R$). It is the difference in the best fit power law index for small planets ($< 3 R_\oplus$) that leads to a fourfold drop in $\eta_\oplus$ between Model\#1 and 4. 

The lower panel of Figure~\ref{fig:bins} further clarifies why excluding  short-period planets results in smaller $\eta_\oplus$.
It shows that the occurrence of small ($1-1.8\,R_\oplus$) planets, calculated by applying the inverse detection efficiency method  over bins with relatively high completeness $> 0.01$\%, 
drops by almost a factor of $\sim$2 from the $\sim 10$\,days bin to the $\sim$30\,days bin. In contrast, the occurrence of large 
($1.8-3.2\,R_\oplus$) planets increases by $\sim 50$\% over the same bins and continues to increase out to 200\,days. Note that the small planets' $\sim 30$\,days bin has an even higher survey completeness that the bin at 120\,days for the $1.8-3.2\,R_\oplus$ planets and does not fall below 0.01\% out to 200\,days. Hence, the drop beyond $\sim 10$\,days in the planet occurrence of $1-1.8\,R_\oplus$ planets is robust. This drop further indicates that the occurrence of small  short-period planets is not representative for the one at longer orbits and such planets should not be included to infer $\eta_\oplus$.  Interestingly, fitting the occurrence of $1-1.8 \, R_\oplus$ planets beyond 12\,days with a single power law gives an index of 0.13, the same as b$_p$ for Model\#4, and an occurrence  of 10\% when integrated over the HZ period of $0.9- 2.2\,P_\oplus$, similar to the low $\eta_\oplus$ values calculated by \texttt{epos} when excluding the population of small short-period planets.

\section{Discussion and outlook} \label{sec:summary}
When considering the sample of most reliable {\it Kepler} candidates ($Robovetter$ score $\ge$0.9), there are no exoplanets detected in the habitable zone of Sun-like stars. As such, extrapolations are necessary to estimate $\eta_\oplus$. Here, we have shown that  extrapolations relying on the population of small ($<$1.8\,R$_\oplus$) short-period ($<25$\,days) planets bias $\eta_\oplus$ to large values, mainly because  the inferred distribution vs planet radius increases toward Earth-sized planets. Excluding this population leads to  a fourfold to eightfold drop in $\eta_\oplus$. The existence of the radius valley \citep{fulton17}, combined with its orbital period  dependence \citep{vaneylen18,martinez2019}, provides strong support that the population of small  short-period planets is contaminated by stripped cores. Therefore,  the occurrence of small short-period planets is not representative for that of planets further away from the star and should not be used to infer the frequency of rocky planets in the HZ. Supporting this statement we have shown that, in the region with high survey completeness and for the most reliable {\it Kepler} candidates, the population of small ($<$1.8\,R$_\oplus$) planets drops beyond 10\,days. How can we then obtain more reliable estimates of  $\eta_\oplus$?

Independent transit or radial velocity detections of small long-period  ($> 100$\,days) {\it Kepler} candidates would provide the most robust approach to measure the frequency of rocky planets close to the HZ. Such followups  would identify the true planets, thus eliminate the use of candidates with a chosen reliability cut, and, being at relatively large orbital periods, reduce the uncertainty when extrapolating into the HZ. While there are a few on-going efforts (e.g., \citealt{bc19}), the faintness of the {\it Kepler} stars, combined with the large orbital period and transit duration of these candidates, makes it unlikely that all of them can be independently confirmed. Statistical validation, which includes ancillary observational evidence, has been also pursued (e.g., \citealt{torres17}) but it cannot be extended to long-period, low signal-to-noise planets \citep{mullally18,bc19}.

Another approach is to quantify the contamination of sub-Neptunes with significantly reduced envelope mass to the population of  small short-period planets. Understanding whether photoevaporation or core-powered mass loss dominate would be an important first step. As core-powered mass loss correlates with the bolometric luminosity of the star, while photoevaporation is driven by high-energy stellar photons, characterizing the radius valley for stars of different spectral types could help distinguishing between these two mechanisms (e.g., \citealt{ginzburg18}). In addition, quantitative comparisons between both models and the {\it Kepler} data, carried out in the same uniform way, would be extremely valuable to test them. Such comparisons could reveal analytic relations for the period-radius distribution under the influence of atmospheric loss that could be  included in \texttt{epos} and used to refine  $\eta_\oplus$ estimates. Alternatively,  observations of young ($\leq$100\,Myr) clusters with TESS could measure the occurrence of primordial  short-period large planets ($1.8-3.2\,R_\oplus$). Subtracting from this population the corresponding old planet population would give the frequency of sub-Neptunes whose atmosphere  has been significantly stripped away from photoevaporation or  planet's cooling. Finally, removing this population from the {\it Kepler} short-period small ($1-1.8\,R_\oplus$) planets would unveil the occurrence of rocky planets that formed like Earth. \\



\vspace{5mm}
\acknowledgments
This material is based upon work supported by the National Aeronautics and Space Administration under Agreement No. NNX15AD94G for the program Earths in Other Solar Systems. The results reported herein benefited from collaborations and/or information exchange within NASA’s Nexus for Exoplanet System Science (NExSS) research coordination network sponsored by NASA’s Science Mission Directorate.

%

\vspace{5mm}
\facilities{{\it Kepler}}


\software{astropy \citep{2013A&A...558A..33A},  \texttt{emcee} \citep{Foreman-Mackey2013}, 
  \texttt{epos} \citep{epos18}, KeplerPORTs  \citep{bc17}.}



\appendix
\section{\texttt{epos} parametric fit}\label{app}
Here, we briefly summarize the key equations used in \texttt{epos} to fit the observed {\it Kepler} exoplanet population. We direct the reader to M18 for a complete description of the code and examples on how to use it\footnote{https://github.com/GijsMulders/epos}.
The planet occurrence rate distribution is described with separable functions in period $P$ and planet radius $R$:
\begin{equation}
{dN\over d\log P \, d\log R} = A f(P) f(R) 
\label{eq:epos}
\end{equation}
where $A$ is a normalization factor and the integral of the function over the simulated planet period and radius range
equals the number of planets per star ($\eta$). In M18, as well as in Model\#1 and \#2 of this letter, the planet orbital period distribution is described by a broken power law:
\begin{equation}
\begin{aligned}
f(P) =
\begin{cases}
\biggl( {P\over P_{\rm br}} \biggr)^{a_P} & \text{if $P < P_{\rm br}$} \\
\biggl( {P\over P_{\rm br}} \biggr)^{b_P} & \text{otherwise} 
\end{cases}
\end{aligned}  
\end{equation}
where the break in orbital period at $\sim 10$\,days for sub-Neptunes was first recognized by \citet{youdin11} and \citet{howard12} and likely reflects the inner edge of protoplanetary disks \citep{mulders15}. When fitting a large range of planet radii the radius distribution also follows a broken power law:
\begin{equation}
\begin{aligned}
f(R) =
\begin{cases}
\biggl( {R\over R_{\rm br}} \biggr)^{a_R} & \text{if $R < R_{\rm br}$} \\
\biggl( {R\over R_{\rm br}} \biggr)^{b_R} & \text{otherwise} 
\end{cases}
\end{aligned}  
\end{equation}
reflecting early findings of a departure from a single power law at $\sim 2\, R_\oplus$ \citep{petigura13}. This type of broken power law in radius is used in Model\#3 through to \#6 in this letter.
\texttt{epos} generates a synthetic planet population via a Monte Carlo approach by random draws from the distributions outlined above. The typical uncertainty in planet radius is included in these Monte Carlo simulations but it is not propagated in the detection  efficiency or vetting.
Uncertainties on the best fit parameters are obtained via a Markov Chain Monte Carlo simulation using
\texttt{emcee} \citep{Foreman-Mackey2013}. For each simulation presented in this study we used 200 walkers for 5000 Monte Carlo iterations and a 1000-step burn-in.

\begin{figure}[h!]
\minipage{0.45\textwidth}
   \includegraphics[width=\linewidth]{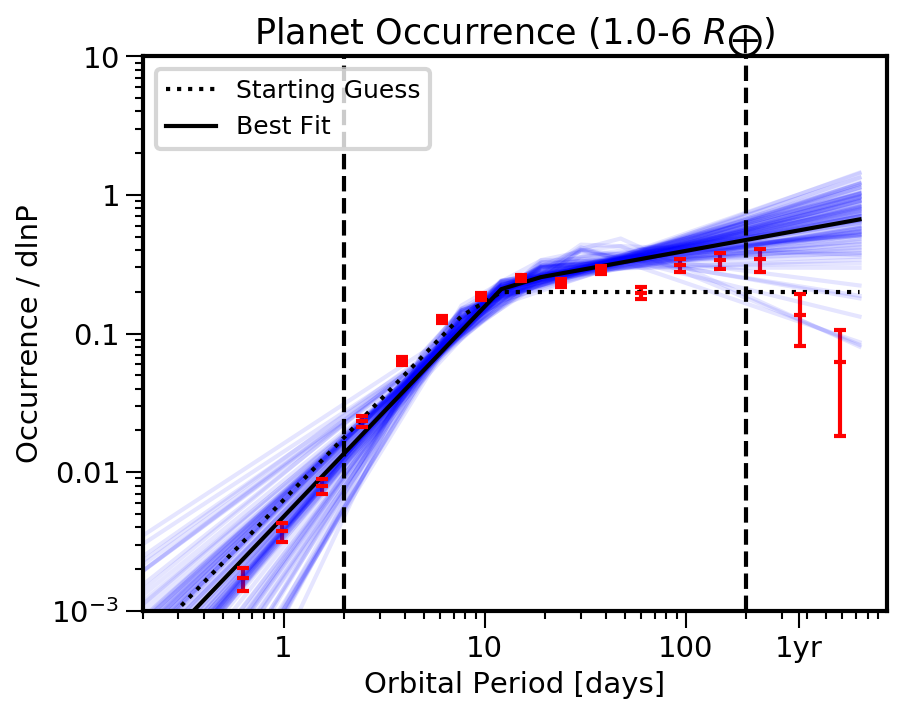}
\endminipage\hfill
\minipage{0.45\textwidth}
   \includegraphics[width=\linewidth]{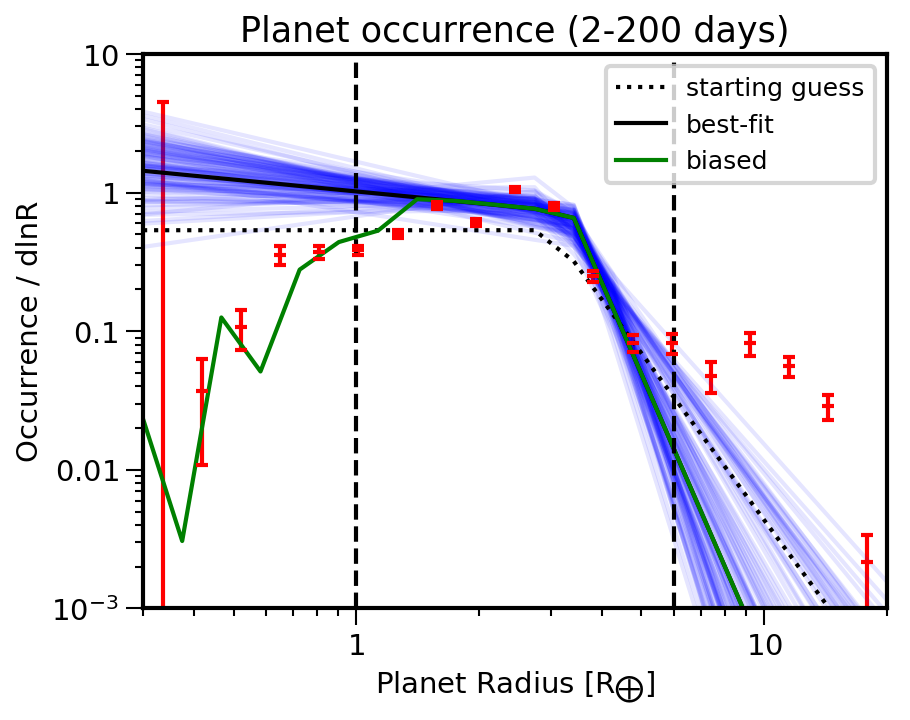}
\endminipage
\caption{ \texttt{epos} posterior orbital period distribution (left) and planet radius distribution (right) for a 2D broken power law as for Model\#1 but with the fit restricted in planet period (2-200\,days) and radius (1-6\,$R_\oplus$). Note that the occurrence rates obtained with the inverse detection method (red points with errorbars) are closer to the best fit than in Figure~\ref{fig:PMulders} as this period and radius range includes fewer bins with low completeness and only partly detected planets.} \label{fig:PMulders_restricted}
\end{figure}

 Figure~\ref{fig:PMulders_restricted} shows the \texttt{epos} posterior orbital period and planet radius distributions for a model analogue to Model\#1 but with the fit restricted in planet period (2-200\,days) and radius (1-6\,$R_\oplus$). This new model results in the same best fit solutions as Model\#1, that is why it is not included in the main text, but illustrates how  the inverse detection efficiency method (red points with errorbars) can underestimate the true occurrence in a bin that includes regions with low survey completeness and planets detected only in part of the bin  (compare the red points in Figure~\ref{fig:PMulders} and Figure~\ref{fig:PMulders_restricted}). The limitations of the inverse detection efficiency method were also pointed out in \citet{fk14}, that is why the forward modeling approach in \texttt{epos} is preferable, especially in regions with few planet detections.

\end{document}